\documentstyle[epsfig]{mn}
\begin{document}
 
\title[Radio observations of Be stars]
{Radio observations of IRAS selected southern hemisphere classical Be stars}

\author[J.S. Clark et al.]{J.S. Clark$^{1,2}$, I.A. Steele$^{3}$ and
R.P. Fender$^{4}$\\
$^1$ Astronomy Centre, University of Sussex, Falmer, Brighton, BN1
9QH, UK\\
$^2$ Department of Physics and Astronomy, University of Southampton,
Southampton, SO17 1BJ, UK \\
$^3$ Astrophysics Research Institute, Liverpool John Moores University,
Liverpool, L3 3AF, UK \\
$^4$ Astronomical Institute `Anton Pannekoek', University of Amsterdam
and Center for High Energy Astrophysics,\\
 Kruislaan 403, 1098 SJ
Amsterdam, Netherlands \\
}
\date{Version 30 March 1998}

\maketitle
\begin{abstract}
We present the first radio observations of a sample of 13 optically and IR
bright southern hemisphere classical Be stars made
from the Australian Telescope Compact Array 
at 3.5cm and 6.3cm simultaneously. One star, $\delta$ Cen was
detected at 3.5cm and a second, $\mu$ Cen was also thought to have been
detected; further observations of this source are required to confirm
this detection. No sources were detected at 6.3cm, although $\delta$
Cen was previously detected at this wavelength by other observers at a
higher flux than our detection limit. The radio observations
show that the spectral energy distribution undergoes a turnover
between the far IR and radio wavelengths, as was seen in previous
studies.  Likewise
we find no simple correlation between far IR and radio flux. 
Lower limits to the outer disc radius were found to be of the order of
few hundred solar radii; of the order of those found
previously by Taylor et al.

\end{abstract}
 
\section{Introduction}

Classical Be stars are defined as non supergiant B stars that have,
or have had Balmer lines in emission. They are further
characterised by the presence of a  continuum excess, 
arising from free free and bound free emission 
from a stellar wind. Comparison of optical
and UV spectra show that two different wind regimes must
coexist. Consequently  a high
velocity component responsible for the high excitation lines visible
in the UV, and a denser component that produces the near IR continuum
excess and the optical emission lines was proposed to explain the
observations. That the denser component was
concentrated in the equatorial plane has long been suspected; 
recent  interferometric data shows that the the envelopes around Be
stars are non spherical
(Dougherty \& Taylor 1992; Quirrenbach et al 1994; Stee et al 1995).
 
However, only a few of the brightest systems are ammeanable to an 
interferometric
approach. Consequently, other approaches to the study of Be star
circumstellar discs have been attempted. One such approach was long
wavelength (mm-radio) flux measurement. When combined with near IR
photometry modeling of the spectral energy distribution
(SED) leads to a profile of the base density and density gradient
(and hence a radial velocity law) within the circumstellar
disc. However, observations of Be stars at mm and radio wavelengths
showed that they were much fainter than implied by a simple
extrapolation of their near IR and IRAS fluxes (e.g. Taylor et al
1990; henceforth Ta90), indicating a change in the ion density
gradient within the disc.
Several explanations were advanced to explain this, 
the most favourable being a change in disc opening angle
or re-acceleration of material at large radii. Other possibilities
exist, such as recombination at large
radii or a truncation of the disc; see  Waters et al. (1991) and Ta90
for a thorough discussion of all the scenarios.

Given the paucity of detections in
the northern hemisphere (only 6 Be stars have been detected), 
and the lack of multiple observations 
it is difficult to quantify the behaviour of the continuum
excess at long wavelengths. Because of this shortfall we made
observations of a sample of 13 bright southern hemisphere Be stars
from the Australian Telescope Compact Array (ATCA) in 1997 April/May
in an attempt to increase the sample size.
 
\section{Observations}
 
Australian Telescope Compact Array (ATCA) observations were made
between 1997 April 30 and 1997 May 4 of a sample of bright Be stars. In
order to obtain a data set consistent with observations of northern
hemisphere Be stars, the same selection criteria used by Ta90 were
adopted (if an extrapolation of the IRAS fluxes revealed a radio flux
of $\sim$1mJy). This procedure resulted in an inadequate number of
stars being selected, the remainder selected on the basis of being
within the declination limits of ATCA and being optically/IR bright
stars listed in the Bright Star Catalogue. 
The list of target stars is reproduced in Table 1;
it consists of targets at a declination of less than
$\sim$0$^{\circ}$. Quoted IRAS fluxes are those given by  Waters et al
(1987). We note that BS7342,  $\theta$ Cir and $\epsilon$  Tuc have
detections listed in the IRAS point source catalogue, however we
only list IRAS fluxes for those stars which satisfy the 
selection criteria listed in Waters et al (1987).
Given the extremely variable nature of
the near IR continuum of Be stars (indeed $\mu$ Cen is also highly variable
in H$\alpha$)  it is not clear that extrapolation
of IRAS data to the present day is valid (either for target selection
or data analysis); therefore caution must be applied in the
interpretation of such data. 

Despite the survey of Ta90 extending to declinations 
of -40$^{\circ}$, only one star $\alpha$ Col 
is included in both surveys. In a second survey of northern
Be stars by Apparao (1990; henceforth Ap90) a further two stars, 
66 Oph and $\chi$ Oph are common to our survey.

The sources were observed with the array in the 1.5A configuration,
giving baselines between 153 and 3000 m, simultaneously at 3.5 \& 6.3
cm. This gave a resolution of 4 arcsecs at 6.3cm and 2 arcsecs at 3.5cm.
Primary flux calibration was achieved using PKS 1934-638. 
Observations typically consisted of 20 min on-target interleaved with
5 min on a nearby unresolved phase calibrator. 
Total observing time on-source was between 1.5-2hrs for each source,
over a period of $\sim 10$ hrs, giving reasonable {\em u-v} coverage.
Data were reduced using the MIRIAD software package.  Radio positions
and strength of each detection were obtained by fitting 2D gaussians
to the radio maps. The typical one sigma rms uncertainty is
$\sim$0.1mJy, 2-3 times more sensitive than previous observations, 
although some fields were
extremely noisy,
resulting in larger upper limits to the radio flux 
(eg the  6.3cm field of  $\pi$ Aqr which contains a very bright
background radio source).

\begin{table*}
\begin{center}
\begin{tabular}{cccccccccc} \hline
Star & HD & SP. Type & S[12] & S[25] & S[60] & 3.5cm Flux Density &
6.3cm Flux Density & Dist. & $L_{radio}$ \\
     &    &          & (Jy)  & (Jy)  & (Jy)  & (mJy) & (mJy) & (Pc)&
($10^{16} erg s^{-1} Hz^{-1}$)  \\ 
\hline
$\alpha$ Col & 37795 & B7IVe & 4.69 & 2.31 & 0.88 &$<$0.09 & $<$0.15 &
85 & $<$0.1 \\
$\delta$ Cen & 105435 & B2IVne & 12.56 & 7.11 & 2.94 & 0.6$\pm$0.03 &
$<$0.2 & 121 & 1.1 \\
$\mu$ Cen & 120324 & B2IVe & 1.42 & 0.63 & - & 0.2$\pm$0.03 & $<$0.12 &
162  & 0.6 \\
$\theta$ Cir$^{\ast}$ & 131492 & B4Vnpe & -  & -  & - &$<$0.16 & $<$0.84
& 256 &  $<$1.3\\
$\mu$ Lup & 135734 & B8Ve & 0.99 & 0.47 & - &$<$0.12 & $<$0.15 &  89 & $<$0.1 \\
$\chi$ Oph & 148184 & B2IVpe & 8.91 & 4.51 & 2.28 & $<$0.3  & $<$0.3 &
150 &  $<$0.8\\
 & 153261$^{\ast}$ & B2IVne & -  & - & -& $<$0.16 & $<$0.25 & 700 & $<$9.4 \\
$\alpha$ Ara & 158427 & B2Vne & 10.12 & 5.07 & 1.68 & $<$0.12 &
$<$0.12 & 74 & $<$0.1\\
66 Oph & 164284 & B2Ve & 2.95 & 1.61 & 0.79 &$<$0.19 & $<$0.09 & 207 & $<$1.0  \\
BS 7342$^{\ast}$ & 181615 & B2Vpe+ & -  & -  & -  & $<$0.15 & $<$0.39
&513 &  $<$4.7 \\
$\pi$ Aqr$^{\ast}$ & 212571 & B1Ve & - & - & - &$<$0.09 & $<$1.2 & 338 &  $<$1.2\\
$\beta$ Psc & 217891 & B6Ve & 0.64 & 0.43 & -  &$<$0.07 & $<$0.15 &
151 &  $<$0.2\\
$\epsilon$ Tuc$^{\ast}$ &  224686 & B9IVe & -  & -  & -  & $<$0.10
& $<$0.12 & 115 &  $<$0.2 \\ 

\hline
\end{tabular}
\caption{Summary of the program stars (distances derived from
Hipparcos archival data, spectral types from the Bright Star Catalogue), 
and observed 3.5cm and 6.3cm fluxes and
upper limits (3 sigma) to fluxes. Stars marked with an asterix were
chosen using optical brightness alone, and not from IRAS continuum
measurements (see text for selection criteria).}

\end{center}
\end{table*}
 
\begin{table*}
\begin{center}
\begin{tabular}{cccc} \hline
Detection & Optical & Radio & Offset (arcsec) \\
\hline
$\delta$ Cen & 12 08 21.50 -50 43 20.7 & 12 08 21.54 -50 43 20.41 &
0.48 \\
$\mu$ Cen & 13 49 36.99 -42 28 25.4 & 13 49 36.84 -42 28 24.74 & 1.81 \\
\hline
\end{tabular}
\caption{Optical (Hipparcos) and radio positions, and positional offsets
 of the two possible radio
detections (all co-ordinates in J2000).}

\end{center}
\end{table*}

\section{Results}

We note that out of a sample of 13 stars we obtained only one 3$\sigma$
detection at 3.5cm; $\delta$ Cen, which Dougherty et al (1998)
confirm ($S_{3.5cm}$=0.67mJy, $S_{6.3cm}$=0.47mJy).
A detection of an unresolved  radio source $\sim$1.8 arcsecs from the
optical position of $\mu$ Cen was made. With a synthesised beam size
of 2.3$\times$1.5 arcsecs this detection is less than a beamsize from
the reported optical position;
for the purpose of this study
we will treat this as a positive detection, although we caution that
further observations are required to confirm this. 
We also note the detection of  an
unresolved radio source of a magnitude consistent with  prior Be star
detections   $\sim$4.5"   south of the optical co-ord of BS7342. 
We exclude this
from further analysis, due to both the distance from the optical
source, and also the peculiar nature of BS7342 (thought to be an
eclipsing  binary
component with an A2 supergiant; Jaschek et al. 1990). We therefore
detect $\sim$15 per cent of the complete target list, and 25 per cent
of the IRAS selected  Be stars (including $\mu$ Cen), compared to
a 30 per cent detection rate by Ta90, and a lack of any detection in
the sample of 18 Be stars by Ap90. 

{\em None} of the sources were detected at 6.3cm; upper limits to the
radio flux are presented in Table 1. Surveys of northern hemisphere
objects reveal that
only two stars, $\psi$ Per and $\beta$ CMi have been detected at 6cm
(Dougherty et al 1991; henceforth D91), with fluxes of 0.23mJy and
0.19mJy respectively. Dougherty et al (1998)
report a detection of $\delta$ Cen at 6.3cm with a flux of
0.47$\pm$0.07mJy, a difference
between the two observations of $\sim$3$\sigma$, suggestive
(although not conclusive) of variability.  We note that the 3.6cm
emission arises from a region interior to that of the 6cm
emission. Therefore, if the changes in the wind propagate {\em outwards}
from the stellar surface the 3.6cm emission would also be expected to vary. 
Of the stars that have been previously observed, $\alpha$
Col was  not detected by either Ta90
or ourselves; the same is true for the two common targets between
this work and Ap90 ($\chi$ Oph and 66 Oph). However, Ap90 failed to
detect any of their sample, including the radio bright Be stars
$\gamma$ Cas, $\psi$ Per, $\zeta$ Tau and $\beta$ Mon.

\section{Discussion}

\begin{figure}
\leavevmode\epsfig{file=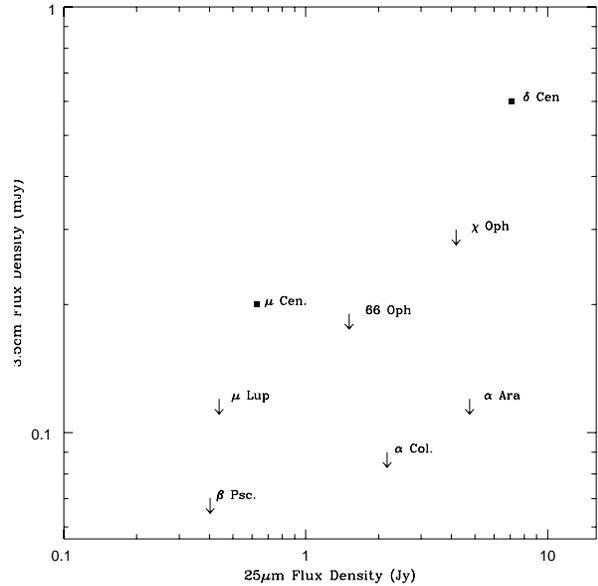,width=8cm,clip=,angle=0,bblly=159,
bbllx=36,bbury=700,bburx=580}
\caption{Detected radio flux and upper limits to flux
($\lambda$=3.5cm) plotted against archival IRAS 25$\mu$m fluxes}
\end{figure}

\begin{table*}
\begin{center}
\begin{tabular}{ccccccc} \hline
Object &  IR-Radio Spectral & 
Radio Spectral & Density Gradient & Minimum Disc &
\multicolumn{2}{c}{Extrapolated 2cm Flux} \\
      &  Index &   Index($\alpha$) & ($\beta$) &  Radius (R$_{\odot}$) &
$\alpha$=observed & $\alpha$=1.4\\
\hline
$\delta$ Cen &  1.29$\pm$0.02 & $>$1.18 &$>$3.1 & 421 & $>$2.1 & 2.4$\pm$0.1 \\
$\mu$ Cen & 1.11$\pm$0.03 & $>$0.87 & $>$2.4  & 326 & $>$1.0 & 1.3$\pm$0.2 \\
\hline
\end{tabular}
\caption{Spectral index, minimum radial density gradient, and minimum
outer disc radius for the 2 radio detections. Also listed are the
extrapolated 2cm fluxes based on the observed lower limits to the
spectral index, and also the canonical value of $\alpha$=1.4 (units of $10^{16} erg s^{-1} Hz^{-1}$) }.
\end{center}
\end{table*}


Like Ta90 we find no strong correlation between radio and far IR
flux (Figure 1). In particular we detect $\mu$ Cen,
despite it having a  relatively low IR flux (it is the faintest Be star
yet to be detected at radio wavelengths).  
However given the length of time that has elapsed
between  the two surveys, it is possible that the far IR flux
has varied during this period; $\mu$ Cen is known to be optically
variable (eg Hanuschik et al 1993). 
If stellar luminosity plays a role in the formation of the circumstellar
envelope, one might expect a correlation between radio emission and
spectral type. Indeed Ta90 point out that the radio fluxes from later
spectral types in their survey are an $\sim$order of magnitude lower
than those of the earlier spectral type. To enable a comparison, we
have extrapolated the fluxes of our detections to 2cm (see Table
3). We present two values here; firstly a lower limit based
on our determination of the spectral index, and
a second estimate based on the assumption of a spectral index of
$\sim$1.4 (suggested by prior work). In both cases we find
that the luminosity is consistent with those measured for the earlier
spectral types in Ta90 ($\sim$10$^{16}$-10$^{17}$ erg s$^-1$ Hz$^-1$),
an $\sim$order of magnitude greater than the detections of later
spectral types reported in Ta90. We further note that the luminosities of 
$\delta$ Cen and $\mu$ Cen 
 differ by 0.5$\pm$0.1 $10^{16} erg s^{-1} Hz^{-1}$ suggesting considerable
variance between the circumstellar discs of these two stars (of the
same spectral type).

Calculation of the 25$\mu$m-3.5cm spectral index are shown in Table
3. For stars for which only upper limits are available we find that
the lower limits to the spectral index are of the order
$\alpha$$\sim$1.2-1.5; indicative of a 
steepening in the spectral energy distribution between IR and radio
wavelengths, as first reported in Taylor et al. (1987). Although two stars
($\theta$ Cir and $\epsilon$ Tuc) do not demonstrate this
behaviour, we note that they have the steepest IR spectral slopes;
therefore we might expect their radio fluxes to lie substantially below
the upper limits determined here. 

This range of spectral indices confirms the
departure of the Be star wind from an isothermal circumstellar disc
with a constant disc opening angle and
wind velocity with respect to radius, and a constant mass loss rate,    
for which one would expect to see a spectral index of +0.6 (D91;
Wright and Barlow 1975). Therefore, as concluded in previous work one
or more of these underlying assumptions must be incorrect. Under the
assumption of a constant opening angle and an isothermal disc it is
possible to relate the spectral index to the density gradient (and
hence radial velocity gradient) within
the circumstellar disc (D91). For a radial density law of the form
${\rho}{\propto}{r^{-{\beta}}}$ we find that the spectral index,
$\alpha$, is related to $\beta$ by (Wright and Barlow 1975)

\begin{equation}
\beta=(6.2-{\alpha})/(4-2{\alpha}) 
\end{equation}

The lower limits to the density gradient implied by the spectral
indices are listed in Table 3. Waters et
al (1987) show that for the majority of Be stars, the value of $\beta$
determined from IRAS measurements is consistently between 2-3. Values
of $\beta$ in excess of 2 are of interest as they imply either an
acceleration of disc material or a non constant disc opening angle (Ta90).
Of the stars detected here, the lower limit for $\mu$ Cen is consistent
with these values, while that of $\delta$ Cen  exceeds
it. Near IR observations of the Be star X Persei also indicate
a high value for $\beta$ (Telting et al 1997). The symmetric 
H$\alpha$ profile argue against an explanation for this value in terms
of a large density/velocity gradient alone, which would lead to a very  
asymmetric profile. Since it is expected that the radio
continuum will be formed at larger radii than 
H$\alpha$ emission, it is possible that the large values of $\beta$ do
represent a real velocity gradient (ie an acceleration) at large radii
 within the disc, as
is suggested by Ta87, and not just a departure from the isothermal
assumption as suggested by Telting et al (1997).
However, the 3.5cm and 6.3cm detections of $\delta$ Cen by Dougherty
et al (1998) are suggestive of a possible change in the spectral index
(to a value close to the canonical value for an isothermal, constant
velocity wind; $\alpha \sim$ 0.6). A  change in the spectral index 
in two stars is also reported in D91; it therefore seems likely that 
changes in the global structure of the
circumstellar envelope and/or velocity profile 
are possible. At present it is impossible
to determine if this is a result of changes in the disc geometry or
velocity structure (ie absence of an accelerating force at large radii
during this time). 

%
%

Under the assumption of an {\em optically thick} disc, we can
derive estimates for the minimum outer disc radius of the Be stars. 
Using the method
detailed in Ta90 we obtained estimates of the minimum radius for an
optically thick spherical wind to produce the observed
fluxes. Since Be star winds  are thought to be both partially optically
thin, and also disc like, this estimate will also serve as a lower
estimate to the outer disc radius (see Table 3). 
The values obtained, of the same order  as
those derived in Ta90 clearly show the disc can extend to large distances from
the star in some systems. In contrast to these values
Cot\'{e} et al (1996)  analyse the near IR SED
of the Be stars  BS7739  and $\eta$ Cen and find that to explain the steep
spectrum the discs must undergo a steepening in density gradient
 at only a few-10R$_{\ast}$. Observations of 4 Herculis (Koubsky et
al. 1997) also show that during the onset of an episode of Be activity
a pseudo photosphere of material was formed in the equatorial regions
of the star, which gradually formed an  extended
envelope around the star. Therefore, it would appear that Be star disc
size and structure can vary widely between individual stars.
Whether such extremes  represent stable 
configurations, or are a result of viewing discs in varying stages of
their evolution is as yet unknown.
 
\section{Conclusions}
As a result of observations of a sample of 13 IR bright southern Be
stars we have identified $\delta$ Cen  as a
radio source, which may also  be variable. 
A second star, $\mu$ Cen, is  also though to be a radio emitter, 
although further observations are needed to confirm
this. This brings the total number of Be stars detected at radio
wavelengths to eight (inclusive of $\mu$ Cen). 
We confirm earlier results of Ta90 that
demonstrate a turnover in the spectral energy distribution between the
far IR and radio wavelengths. We find no compelling evidence for a
direct correlation between stellar or far IR luminosity and  radio
flux. Clearly further observations of Be stars at higher
sensitivities are required before such a correlation can be confirmed
or rejected. Measurement of the minimum outer disc radius reveals that
the circumstellar discs extend to several hundred solar radii, again
reproducing the results of Ta90.

\section*{Acknowledgements}

The Australia Telescope is funded by the Commonwealth of Australia 
for operation as a National Facility managed by CSIRO. This work was
carried out partly with Starlink hardware and software. We wish to
thank the staff at ATCA for their help in obtaining the observations,
and Sean Dougherty and  Rens Waters for many interesting and 
helpful discussions during the preparation of this paper.

\end{document}